\documentclass[11pt,a4paper]{article}

\usepackage{latexsym}
\usepackage{amssymb}
\usepackage{amsmath}
\usepackage[authoryear]{natbib}

\newif\ifpdf
\ifx\pdfoutput\undefined
   \pdffalse
\else
   \pdfoutput=1
   \pdftrue
\fi

\ifpdf
    \usepackage{color}
    \definecolor{myred}{rgb}{0.5,0,0}
    \definecolor{myblue}{rgb}{0,0,0.75}
    \definecolor{mygreen}{rgb}{0,0.5,0}
   \usepackage[pdftex,
               pdftitle={On the monotonicity of default probabilities},
               pdfsubject={Rating systems},
               pdfkeywords={Score function; conditional default probability; Neyman-Pearson test},
               pdfauthor={Dirk Tasche},
               pdfstartview=FitBH,
               breaklinks=true,
               colorlinks=true,
               citecolor=myred,
               linkcolor=myblue,
               urlcolor=mygreen]{hyperref}
\else
   \usepackage{hyperref}
\fi

\newtheorem{theorem}{Theorem}[section]

\newtheorem{proposition}[theorem]{Proposition}
\newtheorem{definition}[theorem]{Definition}

\newtheorem{assumption}[theorem]{Assumption}

\numberwithin{equation}{section}

\begin{document}
\title{Remarks on the monotonicity of default probabilities%
}

\author{
Dirk Tasche\thanks{Deutsche Bundesbank, Postfach 10 06 02, 60006 Frankfurt am
Main, Germany\newline E-mail: dirk.tasche@bundesbank.de}\ %
\thanks{The opinions expressed in this article are those of the
author and do not necessarily reflect views shared by Deutsche Bundesbank.} }

\date{July 23, 2002}
\maketitle

\begin{abstract}
The consultative papers for the Basel~II Accord require rating systems to
provide a ranking of obligors in the sense that the rating categories indicate
the creditworthiness in terms of default probabilities. As a consequence, the
default probabilities ought to present a monotonous function of the ordered
rating categories. This requirement appears quite intuitive. In this paper,
however, we show that the intuition can be founded on mathematical facts. We
prove that, in the closely related context of a continuous score function,
monotonicity of the conditional default probabilities is equivalent to
optimality of the corresponding decision rules in the test-theoretic sense. As
a consequence, the optimality can be checked by inspection of the ordinal
dominance graph (also called Receiver Operating Characteristic curve) of the
score function: it obtains if and only if the curve is concave. We conclude
the paper by exploring the connection between the area under the ordinal
dominance graph and the so-called Information Value which is used by some
vendors of scoring systems.\\[1ex]
\textsc{Keywords:} Conditional default probability, score function, most
powerful test, Information Value, Accuracy Ratio.

\end{abstract}

\section{Introduction}

In its new attempt -- the so-called Basel~II Accord -- to provide quantitative
rules for the capital banks are charged with for their credit risks, the Basel
Committee requires banks to determine \emph{default probabilities} for all
obligors in their credit portfolios. These default probabilities can be
derived from internal rating systems \citep{Basel01}. As a consequence, there
is a growing need to develop internal rating systems in order to meet the
requirements by the Basel~II Accord.

However, the Basel~II Accord does not only allow internal ratings but also
gives rules for properties of the rating systems which will be checked by and
then by the supervisory authorities. Hence, the rating systems used by the
banks have to meet certain quality standards. These standards include the
actual state of the system as well as the process of its development. As a
consequence, a lot of work has been done by several researchers in order to
establish a common standard based on reasonable economic and statistical
assumptions.

Monotonicity of the rating system in the sense that better ratings should
correspond to lower default probabilities is one of the most important
requirements \citep[see e.g.][]{Fritz02, Krahnen01}. In case that the rating
system is based on a score function \citet{Fritz02} even demand monotonicity
at score level. In the paper at hand, we show that this requirement can be
based on a decision-theoretic foundation. We prove that monotonicity of the
conditional default probabilities is equivalent to optimality of the
corresponding decision rules in the test-theoretic sense.

This paper is organized as follows. In Section \ref{sec:assump} we present the
mathematical framework and some basic statistical facts on score functions.
Section \ref{sec:mon} gives in Proposition \ref{pr:main} the main result. In
Section \ref{sec:inf} we discuss the connection between two important summary
statistics for the discriminatory power of score functions: the Information
Value (IV) and the Accuracy Ratio (AR).

\section{Assumptions and basic facts}
\label{sec:assump}

In the sequel we describe the result of the scoring process by a real random
variable (or statistic) $S$ on a probability space $(\Omega, \cal{F},
\mathrm{P})$. A second statistic $T$ (for \underline{t}ype) with values in the
set $\{D, N\}$ ($D$ for \underline{d}efault and $N$ for
\underline{n}on-default) indicates whether the firm which has been scored will
be insolvent or solvent by a previously fixed time horizon. As, however, the
value of $T$ can be observed only with some delay the financial institution
faces the problem to infer its value from the known value of $S$.

In order to describe formally the problem we fix some assumptions and
notations:
\begin{itemize}
  \item Write short $D$ for the event $\{T = D\}$ and $N$ for $\{T = N\}$.
  \item Denote the \emph{overall default probability} by $p$, i.e.\
  $p = \mathrm{P}[D] \in (0,1)$.
  \item Assume that $S$ has a conditional density $f(s|t)$ given $T$, i.e.\
\begin{equation*}
  \mathrm{P}[S \in A \,|\,T = t] \,=\, \int_A f(s|t)\, d s
\end{equation*}
for both $t=D$ and $t=N$ and any Borel subset $A$ of the real line. For the
sake of brevity we write
\begin{equation*}
  f_D(s) = f(s|D)\qquad\text{and}\qquad f_N(s) = f(s|N).
\end{equation*}
\item Given the conditional densities $f_D$ and $f_N$ of $S$ for the two
possible values of $T$, we denote the conditional distribution functions of
$S$ given the values of $T$ by $F_D$ and $F_N$ respectively, i.e.\
\begin{equation}\label{eq:condist}
\begin{split}
  F_D(s)& = \mathrm{P}[S \le s\,|\,D] \,=\,\int_{-\infty}^s f_D(x)\,dx, \\
  F_N(s)& = \mathrm{P}[S \le s\,|\,N] \,=\,\int_{-\infty}^s f_N(x)\,dx.
\end{split}
\end{equation}
\end{itemize}
\begin{assumption}[Smoothness of model]\label{as:1}
The densities $f_D$ and $f_N$ are positive continuous functions in some open
interval $I\subset \mathbb{R}$. For any $x > 0$ the set
\begin{equation*}
  U_x \,=\, \{s \in I:\, x\,f_N(s) = f_D(s)\}
\end{equation*}
has Lebesgue measure 0.
\end{assumption}
Note that under Assumption \ref{as:1} both $F_D$ and $F_N$ are continuously
differentiable and strictly increasing functions. In particular, their inverse
functions $F_D^{-1}$ and $F_N^{-1}$ respectively exist and are uniquely
defined.
\begin{definition}[Likelihood ratio]
We call
\begin{equation}\label{eq:odds}
  L(s)\,=\,\frac{f_D(s)}{f_N(s)},\quad s \in I,
\end{equation}
the \emph{likelihood ratio} at score $s$.
\end{definition}
From Assumption \ref{as:1} follows that the conditional distributions
$\mathrm{P}[L\circ S \in \cdot\,|\,D]$ and $\mathrm{P}[L\circ S \in
\cdot\,|\,N]$ of the likelihood ratio applied to the score statistic given the
type of the firm under consideration are continuous. To see this note that
e.g.\
\begin{equation*}
  \mathrm{P}[L\circ S = x\,|\,D] \,=\, \mathrm{P}[S\in U_x\,|\,D]
   \, =\, \int_{U_x} f_D(s)\, d s\,=\, 0
\end{equation*}
for arbitrary $x > 0$.

A further conclusion from Assumption \ref{as:1} is a well-known formula
\citep[e.g][p.~4]{Kullback59} for the
\emph{conditional default probability} $\mathrm{P}[D\,|\,S = s]$ of a firm
given that its score equals $s$, namely
\begin{equation}\label{eq:condprob}
\mathrm{P}[D\,|\,S = s]\,=\,\frac{p\,f_D(s)}{p\,f_D(s)+(1-p)\,f_N(s)}\,=\,
1-\bigl(1 + {\textstyle\frac p{1-p}}\,L(s)\bigr)^{-1}.
\end{equation}
Since for all $s\in I$ we have $\mathrm{P}[S=s] = 0$, this conditional
probability has to be understood in the non-elementary sense
\citep[cf.][ch.~4]{D96}.

If we assume that insolvent firms in general receive lower scores than solvent
firms, a very simple procedure for a test of the hypothesis $T=N$ against the
alternative $T=D$ is to fix a score level $s_0 \in I$ and to reject $T=N$
whenever $S < s_0$. Of course, in case of higher scores for the insolvent
firms the hypothesis $T=N$ should be rejected whenever $S > s_0$. In the
sequel, we will restrict ourselves to the consideration of the first case
since for most of the results the transfer to the second case is obvious.

The level $s_0$ should be chosen in such a way that a maximal level of the
\emph{Type I error} (i.e.\ to reject $T=N$ although it is true) is
guaranteed. For a given level $u\in (0,1)$ this can be achieved by setting
\begin{equation} \label{eq:size}
s_0 = F_N^{-1}(u).
\end{equation}
The classic way to compare the discriminatory power of different decision
statistics is to fix the Type I error at some level and to look at the size
of the \emph{Type II error} (i.e.\ $T=N$ is not rejected although it is
false). In the context of the test procedure specified by (\ref{eq:size}) the
size of the Type II error is $\mathrm{P}[S \ge s_0\,|\,D]=1-F_D(s_0)$. For
convenience, in the sequel we will call tests of this kind \emph{cut-off}
tests. Of course, it does not matter whether the Type II error is
minimized or $F_D(s_0)$ (often called the \emph{power} of the test) is
maximized over all possible statistics $S$. By convention, we consider here
the maximization variant.

The function $\delta: (0,1) \to [0,1]$ defined by
\begin{equation}\label{eq:dom}
 \delta(u)\,=\, \delta_S(u)\,=\,F_D\bigl(F_N^{-1}(u)\bigr), \qquad u \in (0,1),
\end{equation}
maps every possible level of the Type I error on the power of the
corresponding cut-off test. If for two test statistics $S_1$ and $S_2$, we
have $\delta_{S_1}(u) \ge \delta_{S_2}(u)$ for any $u$ then we know that $S_1$
delivers uniformly more powerful cut-off tests than $S_2$. Plotting the graphs
of $\delta_{S_1}$ and $\delta_{S_2}$ provides a convenient way to check this.

In the literature, the graph of $\delta$ is known as \emph{ordinal dominance}
graph or as \emph{receiver operating characteristic} graph \citep[see
e.g.][]{Bamber75}. There is a well-known connection between the function
$\delta$ and the likelihood ratio defined by (\ref{eq:odds}), namely
\begin{equation}\label{eq:deriv}
  \frac{d\,\delta(u)}{d\,u}\,=\,L\bigl(F_N^{-1}(u)\bigr), \qquad u
  \in (0,1).
\end{equation}
As already noticed by \citeauthor{Bamber75}, from (\ref{eq:deriv}) follows
that $\delta$ is concave in $u$ if and only if the likelihood ratio $L$ is
non-increasing in $s$ (or, by (\ref{eq:condprob}), equivalently the
conditional default probability is non-increasing in $s$). Similarly, $\delta$
is convex in $u$ if and only if $L$ and the conditional default probability
are non-decreasing in $s$.

Note that tests for $T=N$ against $T=D$ need not necessarily be of cut-off
form. In general, it seems reasonable to allow for all tests which are
specified by some \emph{rejection range} $R \subset \mathbb{R}$. Such a test
would reject $T=N$ if $S\in R$. In case of a cut-off test, $R$ is described by
\begin{equation}\label{eq:reject}
  R\,=\,(-\infty, s_0).
\end{equation}
As an example for other sensible forms of the rejection range, consider the
case when insolvent firms tend to receive scores close to 0 whereas solvent
firms achieve negative or positive scores with large absolute values. In this
case the choice $R = (s_l, s_u)$ for some numbers $s_l < 0 < s_u$ appears more
appropriate than (\ref{eq:reject}). As a consequence, comparing this score
function with a score function which assigns low values to insolvent and high
values to solvent firms by means of the function $\delta$ from (\ref{eq:dom})
would give a biased impression.

How can this problem be revealed? In Section~\ref{sec:mon}, we will show that
function $\delta_S$ is concave if and only if the $S$-based most powerful
tests of $T=N$ against $T=D$ are cut-off. Hence, a non-concave $\delta_S$ (and
equivalently non-monotonous conditional default probabilities) would indicate
that the information which is contained in the statistic $S$ is not optimally
exploited.

\section{Monotonous conditional default probabilities and optimal
cut-off tests}
\label{sec:mon}

For a proper formulation of the connection between conditional default
probabilities and optimal cut-off tests we need a bit more of mathematical
notation.
\begin{definition}[Randomized test]\label{de:ran}
Let statistics $S$ and $T$ like in Section~\ref{sec:assump} with values in $I$
and $\{N, D\}$ respectively be given. Then any measurable function $\phi:\,I
\to [0,1]$ is called $S$-based (randomized) \emph{test} for $T=N$ against
$T=D$.\\
An $S$-based test $\phi$ is called \emph{test at level} $\alpha \in
(0,1)$ if\/ $\mathrm{E}[\phi\circ S\,|\,N] \le \alpha$.\\
An $S$-based test $\phi^\ast$ is called \emph{most powerful test at level}
$\alpha \in (0,1)$ for $T=N$ against $T=D$ if it is a test at level $\alpha$
and for any $S$-based tests $\phi$
at level $\alpha$ we have
\begin{equation}\label{eq:most}
  \mathrm{E}[\phi^\ast\circ S\,|\,D]\, \ge\,\mathrm{E}[\phi\circ S\,|\,D].
\end{equation}

\end{definition}
We interpret the value $\phi(s)$, $s\in I$, of a randomized test as the
probability of success which should be applied in an additional Bernoulli
experiment conducted by the user of the test in order to decide whether $T=N$
should be rejected. Hence, if e.g.\ $\phi(s) = 0.6$, the user would throw a
coin with success probability $0.6$ and would reject $T=H$ in case of success
only. If $\phi(s) = 1$, the hypothesis has to be rejected unconditionally. In
case $\phi(s) = 0$ it must not be rejected. Tests at level $\alpha$ are just
those tests which guarantee a maximal level $\alpha$ of the Type I error
(i.e.\ to reject $T=N$ although it is true). Most powerful tests at level
$\alpha$ are those tests which minimize the Type II error size among all
the tests at level $\alpha$.

Randomized tests are not widespread in practice since it appears very strange
to throw a coin in order to decide about the rejection of a hypothesis.
However, the common deterministic tests are a subclass of the randomized tests
(they are just those tests with values in $\{0,1\}$ only), and the notion of
randomized test is quite convenient for the formulation of a complete theory
of test optimality.

Our main result will be based on the Neyman-Pearson Fundamental Lemma. We
quote it here in a form adapted to Assumption \ref{as:1} in order to avoid
some technical difficulties. See \citet[][ch.~2.7]{Witting78} for more general
versions.
\begin{theorem}[Neyman-Pearson Fundamental Lemma]\label{th:Neyman}\ \\
Fix a Type I error level $\alpha\in (0,1)$. Then, under Assumption \ref{as:1}, an $S$-based
test $\phi$ for $T=N$ against $T=D$ is most powerful at level $\alpha$ if and only if for
Lebesgue-almost $s\in I$
\begin{equation}\label{eq:test}
  \phi(s)\,=\,
  \begin{cases}
  1, & \text{if}\ L(s) > L_\alpha\\
  0, & \text{otherwise},
  \end{cases}
\end{equation}
where $L$ is the likelihood ratio defined by (\ref{eq:odds}) and
$L_\alpha$ is any constant such that $\mathrm{P}[L\circ S > L_\alpha] = \alpha$.
\end{theorem}
From Theorem \ref{th:Neyman} we know that there is a deterministic most powerful test
at level $\alpha$, namely the test with rejection range $R = L^{-1}\bigl((L_\alpha, \infty)\bigr)$.

If the likelihood ratio $L$ is non-increasing then the test from
(\ref{eq:test}) has cut-off form as described in Section \ref{sec:assump},
i.e.\ there is a constant $s_\alpha$ such that the rejection range is
$(-\infty, s_\alpha)$. Hence, from the observations in Section
\ref{sec:assump} it is clear that concavity of the ordinal dominance graph
implies that the most powerful $S$-based tests for $T=N$ against $T=D$ have
cut-off form. However, this concavity is not only a sufficient but even a
necessary condition for the most powerful tests to be cut-off.
\begin{proposition}[Optimality of cut-off tests]\label{pr:main}\ \\
Under Assumption \ref{as:1}, there is for every Type I error level $\alpha \in (0,1)$ a most
powerful $S$-based test with rejection range of the form $(-\infty, s_\alpha)$ for some
$s_\alpha \in \mathbb{R}$ if and only if the ordinal dominance function $\delta_S$ as defined by
(\ref{eq:dom}) is concave. Similarly, there is for every Type I error level $\alpha \in (0,1)$ a most
powerful $S$-based test with rejection range of the form $(s_\alpha, \infty)$ for some
$s_\alpha \in \mathbb{R}$ if and only if the ordinal dominance function $\delta_S$ as defined by
(\ref{eq:dom}) is convex.
\end{proposition}
In the Appendix, we provide a proof for the statement that existence of most powerful
cut-off tests as in the first part of Proposition \ref{pr:main} implies concavity of the
ordinal dominance function.

Note the dependence on the underlying statistic $S$ in Proposition \ref{pr:main}. Both the most
powerful tests as well as the ordinal dominance function $\delta_S$ are defined in terms of
$S$ and its conditional distributions. As a consequence, there might be another statistic $S^\ast$ such
that there are $S^\ast$-based tests that are more powerful than the corresponding $S$-based tests
at the same Type I error level. Thus, Proposition \ref{pr:main} gives a statement on the optimal use
of available information when the score function has been fixed. The process of finding an appropriate
score function is not subject of the proposition.

\section{Information value and the area under the ordinal dominance graph}
\label{sec:inf}

In section~\ref{sec:mon}, we have investigated which conclusions can be drawn
from concavity or non-concavity of the ordinal dominance graph. In this
section, we will compare the area under the ordinal dominance graph as a
performance measure for score functions with the so-called Information Value
to be explained below.

The natural logarithm of the likelihood ratio defined by (\ref{eq:odds}) is
sometimes called \emph{weight of evidence} \citep[see][]{Good50}. It is used
by some vendors of scoring systems as a means to detect failures in exploiting
the full information of a statistic. Observe that the usefulness of this
method is underpinned by Proposition \ref{pr:main}.

Of course, the weight of evidence is only a local measure of the information
content of a statistic. \citet[][p.~6]{Kullback59} suggested the
\emph{Information Value} (or \emph{divergence}) as the corresponding global
measure. It is defined by
\begin{equation}\label{eq:IV}
  IV_S\,=\,\int_I\bigl(f_D(s) - f_N(s)\bigr) \log L(s)\,d s\,=\,\mathrm{E}[\log L\circ S\,|\,D] -
\mathrm{E}[\log L\circ S\,|\,N].
\end{equation}
From the first representation in (\ref{eq:IV}) follows that $IV_S$ is always
non-negative and symmetric in $f_N$ and $f_D$. In order to arrive at a
test-theoretic interpretation of the Information Value, observe that it can be
equivalently written as
\begin{equation}\label{eq:IVdiff}
  IV_S \,=\, \int_I \bigl(\mathrm{P}[\log L\circ S \le x\,|\,D] -
  \mathrm{P}[\log L\circ S \le x\,|\,N]\bigr) d x.
\end{equation}
Hence $IV_S$ is just the sum of the signed areas between the graphs of the
distribution functions of $\log L\circ S$ conditional on $T=N$ and $T=D$ respectively.
Define
\begin{equation}\label{eq:ast}
  S^\ast \,=\, \log L\circ S,\quad\text{and}\quad F_N^\ast(s)\,=\,
  \mathrm{P}[S^\ast\le s\,|\,N],\ F_D^\ast(s)\,=\,
  \mathrm{P}[S^\ast\le s\,|\,D],\ s \in I,
\end{equation}
and assume that $F_N^\ast$ is differentiable and strictly increasing such that $(F_N^\ast)^{-1}$
exists (and hence is differentiable, too). With the substitution $u = F_N^\ast(s)$, then (\ref{eq:IVdiff})
can be written as
\begin{equation}\label{eq:dens}
  IV_S \, =\,\int_0^1 \bigl(F_D^\ast\bigl((F_N^\ast)^{-1}(u)\bigr) - u\bigr)
  \frac{d (F_N^\ast)^{-1}}{d u}(u)\, d u.
\end{equation}
Replacing the derivative $\frac{d (F_N^\ast)^{-1}}{d u}(u)$ from
(\ref{eq:dens}) by the constant $1$ yields the integral
\begin{equation}\label{eq:deltaast}
  \int_0^1 \bigl(F_D^\ast\bigl((F_N^\ast)^{-1}(u)\bigr) - u\bigr) d u \, = \,\int_0^1 \bigl(\delta_{S^\ast}(u)
  - u\bigr) d u,
\end{equation}
with $\delta_{S^\ast}$ defined analogously to $\delta_S$ in (\ref{eq:dom}).
Here, the right-hand side of (\ref{eq:deltaast}) measures the area between the
ordinal dominance graph of the statistic $S^\ast$ and the diagonal. Twice this
area is a well-known performance measure for scoring systems -- the so-called
\emph{Accuracy Ratio}. The area plus $1/2$ is just the average power of the
$S^\ast$-based cut-off tests where the average is computed with equal weight
$1$ for all Type~I error levels in $(0,1)$. Recall, however, that in case of
(\ref{eq:deltaast}) the score function $S$ has been replaced by the score
function $S^\ast$.

By (\ref{eq:dens}) and (\ref{eq:deltaast}) we have seen that the concepts of information value and
Accuracy Ratio are quite similar from the computational point of view. Nevertheless, they differ
essentially in two aspects. First, the information value is calculated for the transformed statistic
$S^\ast$. And second, for $IV_S$ the difference between the ordinal dominance graph and the diagonal is
weighted with the derivative of the inverse conditional distribution function of $S^\ast$ given $N$.

Another way to express the similarity between the two concepts is to write the
Accuracy Ratio $AR_S = 2 \int_0^1 \bigl(\delta_S(u) - u\bigr)\,d u$ as
\begin{equation}\label{eq:AR}
AR_S \,=\,2\,\bigl(\mathrm{E}[F_D\circ S\,|\,N] - \mathrm{E}[F_D\circ
S\,|\,D]\bigr).
\end{equation}
Hence, $AR_S$ can be generated from $IV_S$ essentially by substituting $F_D$
for $\log L$ in (\ref{eq:IV}). Both $AR_S$ and $IV_S$ can be interpreted as
difference of the conditional expectations of a transformation of $S$ given
$T=N$ and $T=D$ respectively. However, the transformation by $F_D$ is
monotonous and yields values in a bounded range for $AR_S$ whereas by
Proposition \ref{pr:main} the transformation by $\log L$ is monotonous if and
only there are most powerful cut-off tests. In general, there is no finite
bound for the value of $IV_S$.


\appendix
\section{Appendix}

We sketch here the proof of the necessity part of Proposition \ref{pr:main}, i.e.\
that concavity of the ordinal dominance function $\delta$ is necessary for the existence
of a most powerful cut-off test of $T=N$ against $T=D$ at any level $\alpha \in (0,1)$.

From (\ref{eq:deriv}) we know that, under Assumption \ref{as:1}, concavity of $\delta$ is equivalent
to the likelihood ratio $L$ being non-increasing. Hence, since $L$ is positive and continuous by assumption,
it suffices to show that for any $r > 0$ there is some $l_r\in I$ such that
\begin{equation}\label{eq:show}
  L^{-1}\bigl((r,\infty)\bigr) \,=\,(-\infty, l_r)\cap I.
\end{equation}
Choose an arbitrary $r> 0$ and let $\alpha = \mathrm{P}[L \circ S> r\,|\,D]$.
By Theorem \ref{th:Neyman}, the test``rejection of $T=N$ if $L\circ S > r$" is most powerful
at level $\alpha$. However, by assumption, there is an $s_\alpha \in I$ such that
$\mathrm{P}[S < s_\alpha] \le \alpha$ and that the test ``rejection of $T=N$ if $S <
s_\alpha$" is also most powerful at level $\alpha$. Again by Theorem \ref{th:Neyman},
it follows that the functions
\begin{equation*}
  \tau_1(s)\,=\,
  \begin{cases}
  1, & \text{if}\ L(s) > r\\
  0, & \text{otherwise},
  \end{cases}\quad
  \text{and}\quad
  \tau_2(s)\,=\,
\begin{cases}
  1, & \text{if}\ s < s_\alpha\\
  0, & \text{otherwise},
  \end{cases}
\end{equation*}
are Lebesgue-almost everywhere equal.
As $L$ is continuous by assumption, we obtain (\ref{eq:show}).

\end{document}